# The critical current density in type-II superconducting bulk materials

Ming Ju Chou [1] and Wei Yeu Chen [2,*]

[1] *Department of Physics, National Taiwan Normal University, Taipei 10610, Taiwan, R.O.C.*

[2] *Department of Physics, Tamkang University, Tamsui 25137, Taiwan, R.O.C.*

(March 1, 2012)

## Abstract

The critical current density $J_c$ in type-II conventional and high-$T_c$ superconducting bulk materials is investigated based on the quantum theory for the vortex dynamics. It is shown that for a constant magnetic field, the critical current density $J_c$ decreases weakly with increasing temperature when $T < T_{dp}$ (depinning temperature); when $T_{dp} < T < T_f$ (boundary fluctuation temperature), $J_c$ is power-law-decaying, and when $T > T_f$, $J_c$ decays exponentially; while for a constant temperature, $J_c$ first decreases, then increases after reaching a maximum, and finally decreases again as the magnetic field increases. These results are in agreement with the experiments.

* Corresponding author. Tel.: 886-2-22346169; fax: 886-2-22346169

E-mail: wychen@mail.tku.edu.tw (W.Y. Chen)

# 1. INTRODUCTION

Vortex dynamics in type-II conventional and high-$T_c$ superconductors, especially the critical current density $J_c$ has been investigated intensively [1-46], after Anderson [2] first pointed out that the critical current density was reached when the Lorentz force on vortex lattice was balanced by the pinning force [3-8] due to inhomogeneities [7-10] in the specimen [11, 12]. It is understood that the quenched disorder always destroys the long-range order of the vortex lattice, after which only short-range order, the vortex bundle, remains [2, 14, 17, 18, 22, 23].

In this paper we investigate the critical current density $J_c$ for type-II superconducting bulk materials based on the quantum theory of vortex dynamics we [23] have developed. By applying this theory, we have calculated the eigenmodes of the Hamiltonian for the vortex bundles, taken into account the quantum, random and thermal averages of the square of the fluctuations of the deformation and free displacement operators, the critical current density is studied through the balance of the Lorentz force and collective pinning force of the vortex bundle.

It is shown that the critical current density $J_c$ decreases weakly with increasing temperature for temperature $T$ less than the depinning temperature $T_{dp}$; while $J_c$ is power-law-decaying when the range of temperature is between the depinning temperature $T_{dp}$ and boundary fluctuation temperature $T_f$, and $J_c$ decays exponentially for temperature $T$ greater than the boundary fluctuation temperature $T_f$ for a constant magnetic field. On the other hand, $J_c$ first decreases, then increases after reaching a maximum, and finally decreases again as the magnetic field increases for a constant temperature.

The rest of this paper is organized as follows. In the next section, a mathematical description of the model is presented. In section 3, the critical current density due to the collective pinning is investigated for a constant magnetic field as well as for a constant temperature. Several general and important issues about our theory are discussed in section 4. Finally, the concluding remarks are given in section 5.

# 2. MATHEMATICAL DESCRIPTION OF THE MODEL

Let us consider a type-II high-$T_c$ or conventional superconductor. The full Hamiltonian of the fluctuation for the flux line lattice (FLL) in the $z$-direction is given by [16-18, 22, 23]

$$H = H_f + H_R \tag{1}$$

where $H_f = H_{kin} + H_e$ represents the Hamiltonian for the free modes [16-18, 22, 23], with $H_{kin}$ the kinetic energy [16-18, 22, 23]

$$H_{kin} = \frac{1}{2\rho} \sum_{\vec{K}\mu} P_\mu(\vec{K}) P_\mu(-\vec{K}) \tag{2}$$

$H_e$ the elastic energy [16-18, 22-24],

$$H_e = \frac{1}{2} \sum_{\vec{K}\mu\nu} C_L K_\mu K_\nu S_\mu(\vec{K}) S_\nu(-\vec{K}) + \frac{1}{2} \sum_{\vec{K}\mu} (C_{66} K_\perp^2 + C_{44} K_z^2) S_\mu(\vec{K}) S_\mu(-\vec{K}) \tag{3}$$

and $H_R$ represents the random Hamiltonian, given as [16-18, 22, 23],

$$H_R = \sum_{\vec{K}\mu} f_{R\mu}(\vec{K}) S_\mu(-\vec{K}) \tag{4}$$

where $(\mu,\nu) = (x,y)$, $\rho$ is the effective mass density of the flux line [25], $K_\perp^2 = K_x^2 + K_y^2$, $P_\mu(\vec{K}), S_\mu(\vec{K})$ are the Fourier transformations of the momentum and displacement operators, and $C_L, C_{11}, C_{44}$ and $C_{66}$ are temperature- and $\vec{K}$-dependent bulk modulus, compression modulus, tilt modulus and shear modulus, respectively [7, 8, 26-31]. $\vec{f}_R(\vec{K})$ is the Fourier transformation of the collective pinning force $\vec{f}_R(\vec{r}) = -\vec{\nabla} V_R(\vec{r})$, with $V_R(\vec{r})$ the random potential energy of the collective pinning [32-35], which is the sum of the contributions of defects within a distance $\xi$ of the vortex core position $\vec{r}$, where $\xi$ is the temperature-dependent coherence length. The correlation functions of the collective pinning random force are assumed to be the short-range correlation [23],

$$\overline{\ll f_{R\alpha}(\vec{k}) f_{R\beta}^*(\vec{k}') \gg_{th}} = \beta(T,B) \delta_{\alpha\beta} \delta(\vec{k} - \vec{k}') \tag{5}$$

with $\overline{\ll \gg}_{th}$ are the quantum, thermal, and random averages, and $\beta(T,B)$ is the temperature- and magnetic field-dependent correlation strength. It is the quantum correlation strength $\beta^Q(T,B)$ in quantum limit, while in the classical limit it is the classical correlation strength $\beta^c(T,B)$.

The free modes Hamiltonian $H_f$ can be diagonalized [16-18, 22, 23] as follows:

$$H_f = \sum_{\vec{K}\mu} [N_{\vec{K}\mu} + \frac{1}{2}] \hbar \omega_{\vec{K}\mu} \tag{6}$$

$$N_{\vec{K}\mu} = \alpha_{\vec{K}\mu}^+ \alpha_{\vec{K}\mu} \tag{7}$$

with $\alpha_{\vec{K}\mu}^+$, $\alpha_{\vec{K}\mu}$ are the creation and annihilation operators for the corresponding eigenmodes as

$$\alpha_{\vec{K}\mu}^+ = \frac{1}{\sqrt{2\hbar}} [\frac{-i}{\sqrt{\rho\omega_{\vec{K}\mu}}} P_\mu(\vec{K}) + \sqrt{\rho\omega_{\vec{K}\mu}} S_\mu(-\vec{K})] \tag{8}$$

$$\alpha_{\vec{K}\mu} = \frac{1}{\sqrt{2\hbar}}[\frac{i}{\sqrt{\rho\omega_{\vec{K}\mu}}}P_\mu(-\vec{K}) + \sqrt{\rho\omega_{\vec{K}\mu}}S_\mu(\vec{K})] \tag{9}$$

where $\mu = 1$ presents the component parallel to the $\vec{K}_\perp$ direction, while $\mu = 2$ is perpendicular to the $\vec{K}_\perp$ direction, and the corresponding eigenmodes spectrum are given by [16-18, 22, 23]

$$\omega_{K1} = [\frac{1}{\rho}(C_{11}K_\perp^2 + C_{44}K_z^2)]^{\frac{1}{2}} \tag{10}$$

$$\omega_{K2} = [\frac{1}{\rho}(C_{66}K_\perp^2 + C_{44}K_z^2)]^{\frac{1}{2}} \tag{11}$$

The equation of motion for the displacement operator $S_\mu(\vec{K})$ can be obtained from Eq. (1) as

$$\rho \ddot{S}_\mu(\vec{K}) + C_L(\vec{K} \cdot \vec{S}(\vec{K}))K_\mu + (C_{66}K_\perp^2 + C_{44}K_z^2)S_\mu(\vec{K}) + f_\mu(\vec{K}) = 0 \tag{12}$$

Then the solution to Eq. (12) is expressed as

$$S_\mu(\vec{K}) = S_\mu^f(\vec{K}) + S_\mu^R(\vec{K}) \tag{13}$$

where $S_\mu^R(\vec{K})$ is the random displacement operator owing to the collective pinning of the random function $\vec{f}_R(\vec{K})$, and $S_\mu^f(\vec{K})$ denotes the free displacement operator which is the displacement operator of the free modes. They are obtained as

$$S_\mu^R(\vec{K}) = [(\vec{K} \cdot \vec{f}(\vec{K}))\frac{K_\alpha}{K_\perp^2}] \cdot \frac{1}{C_{11}K_\perp^2 + C_{44}K_z^2} + [f_\alpha(\vec{K}) - (\vec{K} \cdot \vec{f}(\vec{K}))\frac{K_\alpha}{K_\perp^2}] \cdot \frac{1}{C_{66}K_\perp^2 + C_{44}K_z^2} \tag{14}$$

and

$$S_\mu^f(\vec{K}) = \sqrt{\frac{\hbar}{2\rho\omega_{\vec{K}\mu}}}[\alpha_{\vec{K}\mu} + \alpha_{-\vec{K}\mu}^+] \tag{15}$$

respectively.

For type-II superconducting bulk materials, the corresponding random and free displacement correlation functions can be obtained as

$$\overline{<<(\vec{S}^R(\vec{R}) - \vec{S}^R(0))^2>>_{th}} = \int \frac{d^3k}{(2\pi)^3} 2(1 - \cos(\vec{K}_\perp \cdot \vec{R}_\perp + K_Z L)) \cdot \frac{\beta(T,B)}{(C_{66}K_\perp^2 + C_{44}K_z^2)^2}$$

$$= \frac{\beta(T,B)}{2\pi^2 C_{66}\sqrt{C_{44}C_{66}}} \cdot [R_\perp^2 + \frac{a_0^2 L^2}{\lambda^2}]^{\frac{1}{2}} \tag{16}$$

and

$$\overline{<<(\vec{S}^f)^2>>_{th}} = \frac{k_B T}{\pi^2 \sqrt{C_{44} C_{66}}} \cdot \frac{1}{\xi_0} \qquad (17)$$

respectively, once again, $\overline{<<\;>>}_{th}$ are the quantum, random and thermal averages, $\beta(T,B)$ is the temperature- and magnetic field-dependent correlation strength, $\vec{R} = (\vec{R}_\perp, L\vec{e}_z)$, with $|\vec{R}_\perp|$ and $L$ are the transversal and longitudinal sizes of the vortex bundle, $\lambda$ is the temperature-dependent penetration depth, the lattice constant $a_0 = (2\Phi_0 / B\sqrt{3})^{1/2}$, $\Phi_0$ is the unit flux, B is the applied magnetic field, $\xi_0$ is the coherence length at zero temperature and $k_B$ is the Boltzmann constant. In deriving the above equations, we have taken into account the fact that $C_{11} >> C_{66}$, performed the average over $\varphi$, which is the angle between $\vec{K}_\perp$ and $\vec{R}$, and applied the cutoff values for small $k$ as $k_s = 2(R_\perp^2 + a_0^2 L^2 / \lambda^2)^{-1/2}$ and for large $k$ as $k_L = 1/\xi_0$.

## 3. CRITICAL CURRENT DENSITY $J_c$

In this section we would like to investigate the critical current density $J_c$ in type-II superconducting bulk materials by considering the balance of the Lorentz force and collective pinning force, $J_c$ can be obtained as [22, 23]

$$J_c = \frac{1}{B}[\frac{\beta(T,B)}{\pi R_\perp^2 L}]^{\frac{1}{2}} \qquad (18)$$

where B is the applied magnetic field, $R_\perp$ and L are the transverse and longitudinal sizes of the vortex bundles, which are determined by the relation

$$\overline{<<(\vec{S}^R(\vec{R}) - \vec{S}^R(0))^2>>_{th}} = \eta^2 \qquad (19)$$

where $\eta$ stands for the collective pinning force range. Let us define the following characteristic temperatures [22] for the upcoming calculations and discussions of the critical current density $J_c$: the division temperature $T_D$, where for $T < T_D$ only the quantum statistics is applicable, while in the case of $T > T_D$ the classical statistics can also be used. The depinning temperature $T_{dp}$, which is defined by the condition

$$\overline{<<(\vec{S}^f)^2>>_{th}} \approx \xi^2 \qquad (20)$$

the boundary fluctuation temperature $T_f$, which is determined by the relation

$$\overline{<<(\vec{S}^f)^2>>_{th}} \approx \chi a_0^2 \qquad (21)$$

with $\chi$ is a dimensionless constant. The collective pinning force range $\eta$ can now be identified for different temperature regimes as

$$\eta \approx \xi \qquad , \quad for \quad T < T_{dp}$$
$$\approx \sqrt{\overline{<<(\vec{S}^f)^2>>_{th}}} \quad , \quad for \quad T > T_{dp} \qquad (22)$$

where the quantum, random and thermal averages of the free displacement correlation function $\overline{<<(\vec{S}^f)^2>>_{th}}$ is given by Eq. (17). The critical current density $J_c$ for type-II bulk materials can now be obtained as follows.

First for temperature $T < T_D$, inserting Eqs. (19) and (22) into (18), the critical current density $J_c$ can be evaluated as

$$J_c = \sqrt{\frac{a_0}{\lambda}} \cdot \frac{\beta^Q(T,B)^2}{2B\sqrt{2\pi}(\pi\xi)^3 C_{66}^{3/2}(C_{66}C_{44})^{3/4}} \qquad (23)$$

where $\beta^Q(T,B)$ is the temperature-and magnetic field-dependent quantum correlation strength that we have discussed earlier in section 2.

For temperature in the interval $T_D < T < T_{dp}$, we have

$$J_c = \sqrt{\frac{a_0}{\lambda}} [\frac{\beta^C(T,B)^2}{2B\sqrt{2\pi}(\pi\xi)^3 C_{66}^{3/2}(C_{66}C_{44})^{3/4}}] \qquad (24)$$

and

$$J_c = \sqrt{\frac{a_0}{\lambda}} [\frac{\beta^C(T,B)^2}{B\sqrt{\pi}(\frac{2k_B T C_{66}}{\xi_0})^{3/2}}] \qquad (25)$$

for $T_{dp} < T < T_f$. Finally, for $T > T_f$, the charge-density-wave type [36-38] pinning regime, we arrive at

$$J_c = \frac{\beta^C(T,B)^{\frac{1}{2}}}{B[\pi R_f^2 L_f \exp(\frac{(3\sqrt{2})k_B T}{a_0 \pi \sqrt{C_{44}C_{66}}} \cdot \frac{1}{\xi_0})]^{1/2}} \qquad (26)$$

where $\beta^C(T,B)$ is the temperature- and magnetic field-dependent classical correlation strength, and $R_f$ and $L_f$ are the transverse and longitudinal bundle sizes at temperature $T = T_f$, respectively.

The above calculated results show that, for $T < T_D$, $J_c$ depends on temperature weakly through the parameters of quantum correlation strength $\beta^Q(T,B)$, coherence length $\xi$, penetration depth $\lambda$, shear modulus $C_{66}$ and tilt modulus $C_{44}$ implicitly, for $T_D < T < T_{dp}$, $J_c$ decreases weakly with increasing temperature through the parameters of classical correlation strength $\beta^C(T,B)$, coherence length $\xi$, penetration depth $\lambda$, shear modulus $C_{66}$ and tilt modulus $C_{44}$ implicitly. For $T_{dp} < T < T_f$, $J_c$ decreases in the form of power law, while for $T > T_f$, $J_c$ decays exponentially with increasing temperature for a constant magnetic field, when $T_{dp} < T < T_f$ (boundary fluctuation temperature), $J_c$ is power-law-decaying, while for $T > T_f$, $J_c$ decays exponentially for a constant magnetic field; on the other hand, $J_c$ first decreases, then increases after reaching a maximum, and finally decreases again as the magnetic field increases for a constant temperature.

### 3.1. Critical Current Density for Constant Applied Magnetic Field

First we would like to investigate the critical current density $J_c$ for constant magnetic field. It is understood that $\beta^C(T,B)$ decreases with increasing temperature for a constant magnetic field B due to the reduction of condensation energy. As we have mentioned above the calculations shown that, for $T < T_{dp}$ the depinning temperature, $J_c$ decreases weakly for both quantum and classical limits, for $T_{dp} < T < T_f$, $J_c$ decreases in the form of power law, finally for $T > T_f$, $J_c$ decays exponentially with increasing temperature. It is interesting to make numerical estimates of the characteristic temperatures $T_D$, $T_{dp}$, $T_f$ and the critical current density $J_c$ as a function of $T/T_c$ for a constant applied magnetic field. We obtained $T_D \approx 10^{-2} K$, $T_{dp} \approx 18.23 K$, $T_f \approx 72 K$, $J_c(10^{-4}) \approx 2.2 \times 10^9 A/m^2$, $J_c(0.192) \approx 2 \times 10^9 A/m^2$, $J_c(0.4) \approx 1.41 \times 10^9 A/m^2$, $J_c(0.758) \approx 3.163 \times 10^8 A/m^2$, $J_c(0.9) \approx 10^6 A/m^2$, and $J_c(1) = 0$. These values are in agreement with the experimental results for $YBa_2Cu_3O_{7-\delta}$ superconducting bulk material [40]. In obtaining the above results the following approximate data have been employed [12]: $B = 1T$, $\xi_0 \approx 1 \times 10^{-9} m$, $\lambda_0 \approx 2 \times 10^{-7} m$, $T_c = 95 K$, $\chi \approx 1.8 \times 10^{-3}$, $\beta^C(T_{dP} = 18.23; B=1) \approx 2.429 \times 10^{-3} N^2/m^3$, $\beta^C(T = 38; B=1) \approx 1.013 \times 10^{-2} N^2/m^3$, $\beta^C(T_f = 72; B=1) \approx 5.06 \times 10^{-3} N^2/m^3$, $\beta^C(T = 85.5; B=1) \approx 6.183 \times 10^{-5} N^2/m^3$, $\beta^Q(T_D = 10^{-2}; B=1) \approx 8.124 \times 10^{-3} N^2/m^3$.

### 3.2. Critical Current Density for Constant Temperature

In this subsection we would like to study the critical current density $J_c$ for constant temperature. It is understood that when the applied magnetic field is above the irreversible line or depinning line, the pins become ineffective, and the vortices can move freely, in other words, $J_c = 0$. for $B \geq B_{irr}$, where $B_{irr}$ is the irreversible field [12]. However, for $B < B_{irr}$, $\beta^C(T,B)$ increases with increasing magnetic field, after reaching a maximum value, then decreases again for constant temperature due to the competition between the reduction of condensation energy and the increase of pinning. The numerical calculations of critical current density $J_c$ as function of $B/B_{irr}$ are also obtained as follows: $J_c(0.01) \approx 5.012 \times 10^8 A/m^2$, $J_c(0.1) \approx 3.17 \times 10^8 A/m^2$, $J_c(0.2) \approx 7.95 \times 10^8 A/m^2$, $J_c(0.3) \approx 10^9 \, A/m^2$, $J_c(0.4) \approx 8.92 \times 10^8 \, A/m^2$, $J_c(0.5) \approx 7.08 \times 10^8 \, A/m^2$, $J_c(0.7) \approx 2.51 \times 10^8 \, A/m^2$, $J_c(0.8) \approx 7.08 \times 10^7 \, A/m^2$, $J_c(1) = 0$. These values are in agreement with the experimental results for $YBa_2Cu_3O_{7-\delta}$ superconducting bulk material [40]. In obtaining the above results the following approximate data have been employed [12]: $T = 72\,K$, $B_{irr} = 10\,T$, $\xi_0 \approx 1 \times 10^{-9}\,m$, $\lambda_0 \approx 2 \times 10^{-7}\,m$, $T_c = 95\,K$, $\chi \approx 1.8 \times 10^{-3}$, $\beta^C(T=72; B=0.1) \approx 7.106 \times 10^{-4}\, N^2/m^3$, $\beta^C(T=72; B=1) \approx 5.06 \times 10^{-3}\, N^2/m^3$, $\beta^C(T=72; B=2) \approx 2.082 \times 10^{-2}\, N^2/m^3$, $\beta^C(T=72; B=3) \approx 4.078 \times 10^{-2}\, N^2/m^3$, $\beta^C(T=72; B=4) \approx 5.719 \times 10^{-2}\, N^2/m^3$, $\beta^C(T=72; B=5) \approx 6.925 \times 10^{-2}\, N^2/m^3$, $\beta^C(T=72; B=7) \approx 1.165 \times 10^{-1}\, N^2/m^3$, $\beta^C(T=72; B=8) \approx 4.179 \times 10^{-2}\, N^2/m^3$.

## 4. DISCUSSION

In this section we would like to make several general and important points about our calculations. First of all, the critical current density $J_c$ in type-II superconducting bulk materials for constant magnetic field as well as for constant temperature has been calculated based on the quantum theory of vortex dynamics we [23] have developed.

Secondly, the model of our calculations is very general it can be used for conventional and high-$T_c$ superconductors. Although their mechanisms and the method of pairing are different, these only affect the structure of the vortex lattice and are a minute effect in our calculations.

Thirdly, the coherence length of the superconductor plays a very important role, since it sets the smallest length scale in our theory that can be seen from the large-$k$ cutoff in the $k$-integration.

Finally, it is worthwhile to know that we only discuss the systems that are in thermodynamic equilibrium with the environment any time dependent behavior shall not be discussed.

## 5. CONCLUSION

The quantum, thermal and random fluctuations of the free displacement as well as the random displacement operator of the flux line lattice and critical current density $J_c$ as a function of temperature as well as applied magnetic field in type-II conventional and high-$T_c$ superconducting bulk materials are investigated based on the quantum model for vortex dynamics that we have developed in both quantum and classical regimes. It is shown that for a constant magnetic field, the critical current density $J_c$ decreases weakly with increasing temperature when $T < T_{dp}$ (depinning temperature); when $T_{dp} < T < T_f$ (boundary fluctuation temperature), $J_c$ is power-law-decaying, and when $T > T_f$, $J_c$ decays exponentially; while for a constant temperature, $J_c$ first decreases, then increases after reaching a maximum, and finally decreases again as the magnetic field increases. These results are in agreement with the experiments.

# REFERENCES

[1] Abrikosov A A. On the magnetic properties of superconductors of the second group. JETP 1957; 5: 1174-82.
[2] Anderson P W. Theory of flux creep in hard superconductors. Phys Rev Lett 1962; 9: 309-11.
[3] Gorter CJ. Note on the superconductivity of alloys. Phys Lett 1962; 1: 69-70.
[4] Gorter CJ. On the partial persistence of superconductivity at very high magnetic fields and current densities. Phys Lett 1962; 2, 26-7.
[5] Kim YB, Hempstead CF, Strnad AR. Critical persistent currents in hard superconductors. Phys Rev Lett 1962; 9: 306-9.
[6] Kim YB, Hempstead CF, Strnad AR. Flux creep in hard superconductors. Phys Rev 1963; 131: 2486-95.
[7] Friedel J, De Gennes PG, Matricon J. Nature of the driving force in flux creep phenomena. Appl Phys Lett 1963; 2: 119-21.
[8] Silcox J, Rolins RW. Hysteresis in hard superconductors. Appl Phys Lett 1963; 2: 231-3.
[9] Labusch R. Dislocations in the flux line lattice. Phys Lett 1966; 22: 9-10.
[10] Träuble H, Essmann U. Defects in the flux line lattice of type-II superconductors. Phys Stat Sol 1968; 25: 373-93.
[11] Campbell AM, Evetts JE. Flux vortices and transport current in type-II superconductors. Adv Phys 1972; 21: 199-428 and references therein.
[12] Brandt EH. The flux-line lattice in superconductors. Rep Prog Phys 1995; 58: 1465-594.
[13] Kleiner WM, Roth LM, Autler SH. Bulk solution of Ginzburg-Landau equations for type-II superconductors: upper critical field region. Phys Rev 1964; 133: A1226-7.
[14] Larkin AI, Ovchinnikov Yu N. Pinning in type II superconductors. J Low Temp Phys 1979; 43: 409-28.
[15] Moore MA. Destruction by fluctuations of superconducting long-range order in the Abrikosov flux lattice. Phys Rev B 1989; 39: 136-9.
[16] Chen WY, Chou MJ. Fluctuation-induced attraction of vortices in anisotropic superconductors. Phys Lett A 2001; 280: 371-5.
[17] Chen WY, Chou MJ. Theories of peak effect and anomalous Hall effect for cuprate superconductors. In: Courtlandt K N, Ed. Superconducting cuprates: properties, preparation and applications, New York: Nova Science Publishers 2009; pp. 213-34.
[18] Chen WY, Chou MJ. Theories of peak effect and anomalous Hall effect in superconducting MgB2. In: Suzuki S, Fukuda K, Eds. Magnesium Diboride (MgB2) Superconductor Research, New York: Nova Science Publishers 2009; pp. 153-74.
[19] Kes PH, Tsuei CC. Two-dimensional collective flux pinning, defects, and structural relaxation in amorphous superconducting films. Phys Rev B 1983; 28: 5126.
[20] Feigel'man MV, Vinokur VM. Thermal fluctuations of vortex lines, pinning, and creep in high-$T_c$ superconductors. Phys Rev B 1990; 41: 8986-90.
[21] Vinokur VM, Kes PH, Koshelev AE. Flux pinning and creep in very anisotropic high temperature superconductors. Physica C 1990; 168: 29-39.
[22] Chen WY, Chou MJ, Feng S. The feature of quantum and thermal fluctuations on collective pinning and critical in superconducting film. Phys Lett A 2003; 316: 261-4.
[23] Chen WY, Chou MJ. The quasiorder-disorder phase transition and the peak effect in type-II conventional and high-Tc superconductor. Supercon Sci Tech 2006; 19: 237-41.
[24] Brandt EH. Flux line lattice in high-Tc superconductors: anisotropy, elasticity, fluctuation, thermal depinning, AC penetration and susceptibility. Physica C 1992; 195: 1-27.
[25] Coffey MW, Clem JR. Vortex inertial mass for a discrete type-II superconductor. Phys Rev B 1991; 44: 6903-8.
[26] Labusch R. Elastic constants of flux line lattice in type-II superconductors. Phys Stat Sol 1967; 19: 715-9.
[27] Labusch R. Elastic constants of the fluxoid lattice near the upper critical field. Phys Stat Sol 1969; 32: 439-42.
[28] Brandt EH. Elastic energy of the vortex state in type II superconductors. I. high inductions. J Low Temp Phys 1977; 26: 709-33.
[29] Brandt EH. Elastic energy of the vortex state in type II superconductors. II. low inductions. J Low Temp Phys 1977; 26: 735-53.

[30] Brandt EH. Order parameter and magnetic field of the distorted vortex lattice and their application to flux pinning in type II superconductors. I. parallel flux lines. J Low Temp Phys 1977; 28: 263-89.

[31] Brandt EH. Order parameter and magnetic field of the distorted vortex lattice and their application to flux pinning in type II superconductors. II. curved flux lines. J Low Temp Phys 1977; 28: 291-315.

[32] Chen WY, Chou MJ. The feature of pinning on vortex-antivortex generation in the type-II superconducting film. Phys Lett A 2000; 276: 145-8

[33] Chou MJ, Chen WY. Vortex-pair production in quenched disorder superconducting film. Phys Lett A 2004; 332: 405-11.

[34] Chen WY, Chou MJ, Feng S. Effect of deformations due to collective pinning on vortex-antivortex production in superconducting film. Phys Lett A 2003; 310: 80-4.

[35] Chen WY, Chou MJ, Feng S. The effect of collective pinning on topological order-disorder phase transition in type-II superconducting films. Phys Lett A 2005; 342: 129-33.

[36] Peierls RE. On the theory of the electric and thermal conductivity of metals. Ann Phys Leipzig 1930; 4: 121-48.

[37] Fröhlich H. On the theory of superconductivity: the one-dimensional case. Proc R Soc A 1954; 223: 296-305.

[38] Fogle W, Perlstein H. Semiconductor-to-metal transition in the blue Potassium Molybdenum Bronze, $K_{0.30}MoO_3$; example of a possible excitonic insulator. Phys Rev B 1972; 6: 1402-12.

[39] Anderson PW, Kim YB. Hard superconductivity: theory of the motion of Abrikosov flux lines. Rev Mod Phys 1964; 36: 39-43.

[40] Küpfer H, Wolf T, Lessing C, Zhukov AA, Lançon X, Meier-Hirmer R, Schauer W, Wühl H. Peak effect and its evolution from oxygen deficiency in $YBa_2Cu_3O_{7-\delta}$ single crystals. Phys Rev B 1998; 58: 2886-94.

[41] Chen WY. Theory of thermally activated vortex bundles flow over the directional-dependent potential barriers in type-II superconductors. The Open Supercon J 2010; 2: 12-8.

[42] Yang HC, Wang LM, Horng HE. Characteristics of flux pinning in $YBa_2Cu_3O_y / PrBa_2Cu_3O_y$ superlattices. Phys Rev B 1999; 59: 8956.

[43] Zheng H, Zhu SY, Chen WY. Quantum phonons and the Peierls transition temperature. Phys Rev B 2001; 65: 014304.

[44] Wördenweber R, Kes PH, Tsuei CC. Peak and history effects in two-dimensional collective flux pinning. Phys Rev B 1986; 33: 3172-80.

[45] Chen WY, Chou MJ. Vortex-pair production and nonlinear flux-flow resistivity in superconducting film with randomly distributed weak pinning sites. Phys Lett A 2001; 291: 315-8.

[46] Zehetmayer M, Eisterer M, Jun J, Kazakov SM, Karpinski J, Birajdar B, Eibl O, Weber HW. Fishtail effect in neutron-irradiated superconducting $MgB_2$ single crystals. Phys Rev B 2004; 69: 054510.